\documentclass[12pt]{article}
\newcommand\be{\begin{equation}}
\newcommand\ee{\end{equation}}
\newcommand\ba{\begin{eqnarray}}
\newcommand\ea{\end{eqnarray}}
\newcommand\eq{\begin{equation}}           
\newcommand\en{\end{equation}}

\input epsf
\usepackage{amssymb}
 
\begin{document}
\title{
{\hfill  \small KAIST-TH2003/03  \\}
{\hfill  \small UCB-PTH-03/08  \\ ~\\~\\}
Successful Modular Cosmology}

\author{
Kenji Kadota$^1$ and Ewan D. Stewart$^2$ \\
$^1$ {\em Department of Physics, University of California, Berkeley, CA 94720,
USA} \\
$^2$ {\em Department of Physics, KAIST, Daejeon 305-701, South Korea}
}
\maketitle   

\begin{abstract}
We present a modular cosmology scenario 
where the difficulties encountered in conventional modular cosmology
are solved in a self-consistent manner,
with definite predictions to be tested by observation.
Notably, the difficulty of the dilaton finding its way to a precarious
weak coupling minimum is made irrelevant by having eternal modular inflation
at the vacuum supersymmetry breaking scale after the dilaton is stabilised.
Neither this eternal inflation
nor the subsequent non-slow-roll modular inflation
destabilise the dilaton from its precarious minimum
due to the low energy scale of the inflation
and consequent small back reaction on the dilaton potential.
The observed flat CMB spectrum is obtained from
fluctuations in the angular component of a modulus
near a symmetric point, which are hugely magnified by
the roll down of the modulus to Planckian values,
allowing them to dominate the final curvature perturbation.
We also give precise calculations of
the spectral index and its running.
\end{abstract}

\setcounter{footnote}{0} 
\setcounter{page}{1}
\setcounter{section}{0} \setcounter{subsection}{0}
\setcounter{subsubsection}{0}

\section{Introduction}
\label{intro}
Early Universe cosmology offers one of the few  opportunities to test
the framework of particle theory phenomenology at high energy scales,
in particular string/M-theory phenomenology.
Indeed, moduli are generic in string/M-theory,
and much attention has been drawn to cosmology where moduli
play important roles in the early Universe \cite{mary}.
This has been referred to as `modular cosmology'.
Unfortunately, despite many attempts to find viable cosmology scenarios
associated with moduli,  
they turned out to give us cosmological obstacles
\cite{Polonyi,dineseiberg,brus}
rather than attractive features.

Our main goal in this paper is to present
a successful modular cosmology scenario
where those difficulties due to moduli 
can be solved in a simple and self-consistent manner.
Furthermore, this scenario gives definite predictions
to be tested by forthcoming observations.

The paper is structured as follows.

Section \ref{moti}
gives our motivation for the investigation of modular cosmology
and discusses the cosmological difficulties involving moduli.
Section \ref{flatonclass} outlines how
those difficulties are overcome in our proposed scenario.

Section \ref{eternal} discusses eternal inflation 
\cite{lindebook} at the vacuum supersymmetry breaking scale,
which is crucial in solving the dilaton stabilization problem. 
Section \ref{modinf} is devoted to the non-slow-roll modular inflation
which naturally follows the eternal inflation,
and whose low energy scale is crucial not to disturb the dilaton potential.

Section \ref{conv} gives a detailed analysis of the density perturbations
produced in our modular cosmology,
starting in Section~\ref{review} with a brief review of
the $\delta N$ formalism \cite{sasaki}
necessary to analyse the contributions of the fluctuations
in the multiple light degrees of freedom
to the final curvature perturbation.

Section~\ref{thermalinf} briefly discusses thermal inflation's solution of the Polonyi/moduli problem.

Section~\ref{dis} gives our conclusions. 

Facing the fact that our knowledge of string/M-theory is still too limited
to fully describe the moduli dynamics in early Universe,
we keep our discussion on string/M-theory aspects fairly general
with emphasis on cosmological consequences to be compared with data.
We hope our scenario can open up new possibilities in model building
of modular cosmology which can be tested by observations.

\section{Motivation}

\subsection{Difficulties in Modular Cosmology}
\label{moti}

Moduli are generic predictions of string/M theory,
but pose major cosmological problems.

The stabilization of the dilaton is a serious problem in both
particle physics and cosmology.
In a typical setup, one simply cannot find a minimum
in the weak coupling regime \cite{dineseiberg}.
More complicated models,
such as the racetrack scheme \cite{racetrack},
can realize a minimum at weak coupling
with a moderate amount of fine-tuning
(which is easily anthropically justified if there is no better solution).
However, even if there exists such a minimum,
it has only a shallow barrier separating it from the runaway part of
the dilaton's potential, and a steeply rising potential on the other side.
Thus any typical cosmological history would lead the dilaton to over-shoot
the desired weak coupling minimum and roll off along the runaway
potential towards zero coupling \cite{brus}.

Another serious cosmological problem is the cosmological moduli problem.
Many moduli are expected to have Planckian vacuum expectation values
relative to any symmetric points where they couple to other fields,
giving them gravitationally suppressed couplings to other light fields.
Shifting of their minima due to supersymmetry breaking in the early universe
causes them to dominate the energy density of Universe,
and they survive until or beyond nucleosynthesis
destroying its successful predictions \cite{Polonyi}.
   
Thus, the ubiquitous moduli predicted by string/M-theory 
have given us a cosmological headache
rather than helping us build a desirable cosmology scenario.
It is therefore worth while to seek realistic modular cosmology scenarios
where those serious cosmological problems are solved without fine-tuning 
in a self-consistent manner,
and we address such a possibility in this paper.

\subsection{Successful Modular Cosmology}
\label{flatonclass}

Let us outline the key ingredients in our 
successful modular cosmology scenario.
 
One of the main features in our scenario is
eternal modular inflation at the vacuum supersymmetry breaking scale.
This is important because the infinite expansion of eternal inflation
compensates for any finite unnaturalness in initial conditions.
In particular, it can compensate for the extreme fine tuning
of initial conditions required for the dilaton to find its way
into a precarious weak coupling minimum.
The importance of the low energy scale is that it ensures that
the dilaton does not get disturbed from its minimum after the
eternal inflation ends locally.
If one had eternal inflation at a higher energy scale
than the scale of vacuum supersymmetry breaking,
the backreaction on the dilaton's potential would dominate
over the dilaton's vacuum potential,
and one could never expect the dilaton to find its way into
its desired weak coupling minimum after the eternal inflation ends.
Moreover, if eternal inflation occurs at a higher energy scale
than the vacuum moduli potential,
the moduli will roll down to possibly several vacua including
run-away vacua, also leading to domain wall problems.

Of course, we do not mean that all of eternal inflation takes place
at this low energy scale, just that the last eternal inflation
(from our local point of view) occurred at that low energy scale.
The full picture of eternal inflation would be of eternal inflation
at the various discrete points in field space,
from the Planck density down, corresponding to
suitable false vacua or maxima with rare, but finitely rare,
transitions between these points, and of course continually
generating non-eternally inflating regions like our own 
\cite{old,new,cosmoewan}.

Since the last eternal inflation occurred at
the vacuum supersymmetry breaking scale,
the inflation that produced the observed
approximately scale-invariant density perturbations
must have occurred at that scale or below.
Such inflation emerges naturally as the modulus rolls away
from its eternally inflating maximum.
The scale-invariance is not automatic but can arise naturally
when proper account of the contributions of the fluctuations in
all the relevant light degrees of freedom in the generically
complicated moduli potential is taken into account.
In particular,
fluctuations in angular degrees of freedom are hugely magnified
by the long roll down of the modulus to Planckian values,
allowing them to dominate the final curvature/density perturbation.

Finally, the overabundant moduli are diluted by thermal inflation
which is natural consequence of thermal effects
on unstable supersymmetric flat directions \cite{lythewan}.

We now look into the details of our scenario in the following sections.

\newpage

\section{Eternal and Post-Eternal Modular Inflation}
\label{eternalmod}
\subsection{Eternal Modular Inflation}
\label{eternal}
The modulus potential has the general form 
\be
V(\Phi) = M_\mathrm{s}^4 \, \mathcal{F}(\Phi/M_\mathrm{Pl}) ~,
\label{moduluspot}
\ee
where $M_\mathrm{s}$ is the scale of supersymmetry breaking
and $M_\mathrm{Pl} \equiv 1/\sqrt{8\pi G}$.
$\Phi$ is the scalar component of a modulus chiral superfield
and $\mathcal{F}$ is a function 
with dimensionless coefficients of order unity.
For most of the rest of the paper, we set $M_\mathrm{Pl} = 1$.
In our vacuum, for the case of gravity mediated SUSY breaking, 
$M_\mathrm{s} \sim 10^{10}-10^{11}$ GeV, in order to get
supersymmetry breaking masses
$m_\mathrm{s} \sim M_\mathrm{s}^2 \sim 10^2-10^3$ GeV.
In our regime of interest, these will be the relevant scales.
\footnote{We expect our scenario to work equally well in the case of
gauge mediated supersymmetry breaking,
but leave proper investigation to \cite{ewankenji}.}

Near a maximum, this potential will have the form
\be
V(\phi) = V_0 - \frac{1}{2} m_\phi^2 \phi^2 + \ldots~,
\ee
where $V_0 \sim M_\mathrm{s}^4$ and $m_\phi \sim M_\mathrm{s}^2$,
and we have simplified to a single real direction $\phi \in \Phi$.
The equation of motion is
\be
\label{psimodular}
\ddot{\phi} + 3H \dot{\phi} - m_\phi^2 \phi = 0
\ee
which, in the approximation $3H^2 \simeq V_0$, has growing solution
\be
\phi \propto e^{\alpha H t}
\propto a^{\alpha}~,
\ee
where
\be
\alpha = \frac{3}{2}
\left( \sqrt{ 1 + \frac{4 m_\phi^2}{3 V_0} }\, - 1 \right)~.
\label{inflatoneqn}
\ee

Classically, one will get eternal inflation at the maximum
as the field never rolls off.
Quantum mechanically, one will have a distribution of field values
which will scale like $a^{\alpha}$.
Thus, the probability of having a field value $|\phi| < \varepsilon$
for some small $\varepsilon$ will go like
\be
P(|\phi|<\varepsilon) \propto \frac{1}{\sqrt{\langle\phi^2\rangle}\,}
\propto a^{-\alpha}~,
\ee
while the physical volume of space scales as $a^3$.
Thus if $\alpha < 3$ or
\be
m_\phi^2 < 6 V_0~,
\ee
one will have eternal inflation at the maximum
\cite{cosmoewan,pusanlecture}.

As we discussed in Section~\ref{flatonclass},
eternal inflation at this low energy scale plays a crucial role
for stabilization of the dilaton in our scenario.

\subsection{Post-Eternal Modular Inflation}
\label{modinf}

Once the modulus locally rolls away from the maximum,
the eternal inflation will end locally,
but one will still get some more inflation as it rolls towards
its minimum at Planckian values.
This inflation is important because fluctuations produced during
the eternal inflation are large leading to an inhomogeneous universe.
The spectrum of perturbations produced by the modulus as it rolls
away from the maximum is \cite{more}
\be\label{Pnat}
P^\frac{1}{2}(k) = 2^\alpha
\frac{\Gamma\left(\alpha+\frac{3}{2}\right)}{\Gamma\left(\frac{3}{2}\right)}
\frac{H_\star}{2\pi\alpha\phi_\star}
\left(\frac{k}{a_\star H_\star}\right)^{-\alpha}~,
\ee
where subscript $\star$ denotes an appropriate evaluation point
\footnote{One would usually choose some time around horizon crossing
as the appropriate evaluation point, but this is not necessary here.
In fact, the evaluation point can be any time during inflation
when the scaling of $\phi$, $\phi \propto a^\alpha$,
and the constancy of $H$ ensure that Eq.(8) is independent of
the time chosen to evaluate it.}.
For our values of $H \sim 10^{-16}$ and $\alpha \sim 1$, 
this drops below the COBE normalisation for
\be
\phi > \phi_\mathrm{COBE} \sim 10^{-12}~.
\ee
Thus to make the observable universe sufficiently homogeneous
we require that observable scales leave the horizon
some time after $\phi = \phi_\mathrm{COBE}$.
The observationally relevant number of $e$-folds is
the number of $e$-folds from this point
until inflation ends at $\phi \sim 1$,
which is
\be
N = \frac{1}{\alpha} \ln\left(\frac{1}{\phi_\mathrm{COBE}}\right)
\simeq \frac{28}{\alpha}~.
\ee
Recalling that, for inflation at $V^{1/4} \sim 10^{10}-10^{11}$ GeV
and assuming radiation domination after the inflation,
the observable universe leaves the horizon around 45 $e$-folds
before the end of inflation \cite{liddle}, and taking account of
an additional $\sim 10$ $e$-folds due to thermal inflation,
we see that for $\alpha \sim 1$ we have enough $e$-folds of inflation
to make the observable universe sufficiently homogeneous.

\section{Density Perturbations}
\label{conv}

Beyond making the universe homogeneous, we also need to produce
the observed approximately scale-invariant spectrum of density perturbations.
The previous simplified model fails to do this for the usual reason:
the supersymmetry breaking effect of the inflationary potential energy density
induces a mass squared $m_\phi^2 \sim V_0$
and hence we have $\alpha \sim 1$ and a spectrum far from scale invariant
\cite{copeland,iss}.

However, to properly calculate the spectrum of density perturbations
produced during modular inflation, we need to take proper account of the
contributions of fluctuations in all the relevant light degrees of freedom
to the final curvature/density perturbation,
and not just use an over-simplified toy model.
We start with a brief review of the formalism for this \cite{sasaki}.

\subsection{$\delta N$ Formalism}
\label{review}

The $\delta N$ formalism \cite{sasaki} relates 
the final curvature perturbation to
the perturbation in the number of $e$-folds of expansion
from an initial spatially flat hypersurface
to a final comoving, or equivalently constant energy density, hypersurface
\be
\mathcal{R}_\mathrm{c} = \delta N~.
\ee

For illustration, let us consider single field dynamics first.
In this case, the field trajectory is unique and
therefore any perturbation just corresponds to a shift along the trajectory,
which can be compensated by a local time shift, and hence
leads to a curvature perturbation depending only on local quantities.
If there exist multiple fields, however, there is
an infinite family of field trajectories in the multiple-field space
and each of the trajectories can follow  different equations of state.
Thus perturbations which shift from one trajectory to another
can change the equation of state history and hence the
number of $e$-folds of expansion
\be
\label{Ne}
N = \int H dt = - \int H \frac{d \rho}{\dot{\rho}}
= \int \frac{d\ln\rho}{3(1+w)}
\ee
in a very non-local way.
Here $w \equiv p/\rho$.
The final curvature perturbation is achieved when the multiple field
dynamics disappears and the trajectories coalesce,
leading to constant $\mathcal{R}_\mathrm{c}$ on superhorizon scales.

To apply this formalism, we simply write
\footnote{In general one has
$\delta N = \frac{\partial N}{\partial \phi_i} \, \delta\phi_i
+ \frac{\partial N}{\partial \dot\phi_i} \, \delta\dot\phi_i$
but usually one can pick out a growing mode for $\delta\phi_i$
in which case $\delta\phi_i$ and $\delta\dot\phi_i$ are no longer
independent.}
\be\label{Ntheta}
\mathcal{R}_c = \delta N
= \frac{\partial N}{\partial \phi_i} \, \delta\phi_i~,
\ee
where $\phi_i$ represents each dynamical degree of freedom.
Note that, in the above expression,
$\partial N/\partial \phi_i$ is a transfer function
depending on background fields
while $\delta \phi_i$ is the perturbation on the initial flat hypersurface,
and they can be calculated separately.

In the following sections we apply the $\delta N$ formalism
to calculate the final curvature perturbation in three concrete toy models
properly taking into account all relevant degrees of freedom.
\footnote{In this paper we mostly apply the $\delta N$ formalism to
dynamics towards the end of inflation. However, it also encompasses cases,
such as curvaton scenarios \cite{curvaton}, where the multiple degrees
of freedom survive long after inflation.}

\subsection{Simple Single Modulus}
As an introduction, let us discuss the simplest case
of a single complex modulus
\be
\Phi = \frac{1}{\sqrt{2}\,} \phi e^{i \theta}~.
\ee
Eq.~(\ref{Ntheta}) gives
\be
\mathcal{R}_\mathrm{c} = \delta N
= \frac{\partial N}{\partial \phi} \, \delta\phi
+ \frac{\partial N}{\partial \theta} \, \delta\theta ~.
\ee
We assume that $\Phi=0$ is a symmetric point so that
the potential near the maximum has the form
\be
V = V_0 - m_\phi^2 |\Phi|^2 + \ldots ~.
\ee
However, now that we are considering multiple real fields,
just knowing the local potential is not good enough.
The full modulus potential will be completely model dependent
but a simple toy model will be sufficient for our purposes.
We take
\be
V = V_0 - m_\phi^2 |\Phi|^2
+ \frac{1}{3} m_\phi^2 \left( \Phi^3 + \mbox{h.c.} \right)
+ m_\phi^2 |\Phi|^{4}
\label{toy}
\ee
with $V_0 = 2 m_\phi^2 / 3$ for vanishing vacuum energy.

The contribution to the final curvature perturbation
from the radial component is
\be
\label{rad2}
\frac{\partial N}{\partial \phi} \, \delta\phi 
= - \frac{H}{\dot{\phi}} \, \delta\phi
\sim \frac{H}{\phi}~,
\ee
and was given precisely in Eq.~(\ref{Pnat}).
The contribution to the final curvature perturbation
from the angular component is
\be\label{ang2}
\frac{\partial N}{\partial \theta} \, \delta\theta 
\sim \frac{\partial N}{\partial \theta} \frac{H}{\phi}~.
\ee
To evaluate this we need to determine
$\partial N/\partial \theta$.

The analytical estimate of $\partial N/\partial \theta$ in our scenario
is non-trivial because the approximate U(1) symmetry at small $\phi$
is heavily broken at large $\phi$.
We hence give numerical analysis to determine $\partial N/\partial \theta$.
As our initial conditions, we take a small value of $\phi$,
where the potential is still angularly symmetric
so that the exact value of $\phi$ does not matter,
and various values of $\theta$.
Some field trajectories are shown in Figure~\ref{fig1}.

{
\begin{figure}[ht]
\epsfxsize= 
3.9in
\begin{center}    
\leavevmode       
\epsffile{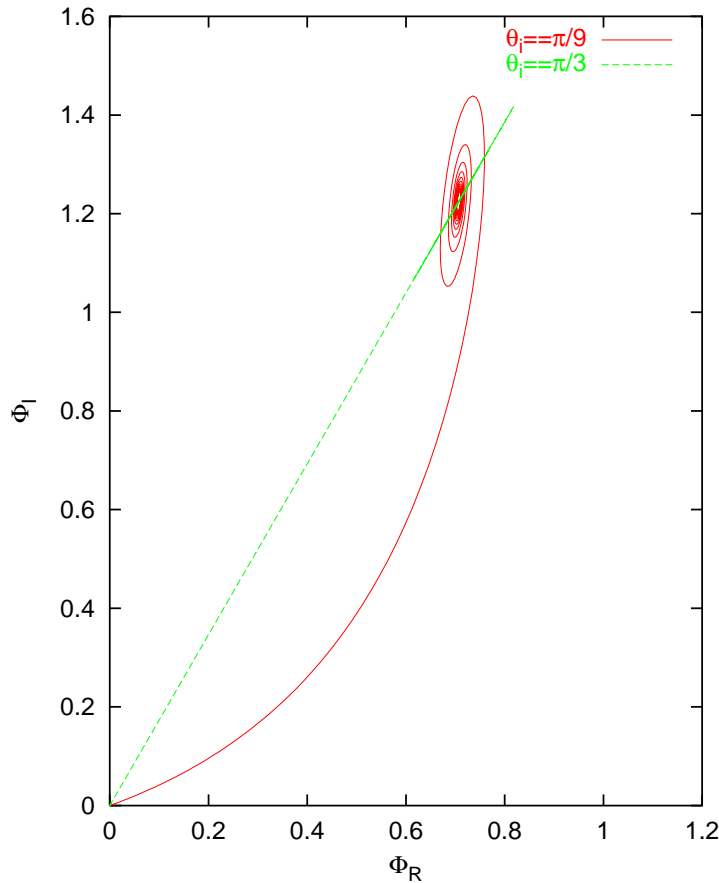}
\end{center}        
\caption{Trajectory of
$\Phi = (\Phi_\mathrm{R} + i \Phi_\mathrm{I}) / \sqrt{2}$
for initial angles $\pi/9$ and $\pi/3$.} 
\label{fig1}
\end{figure}
}

We can see that the initial angle $\theta_i$ greatly affects
the form of the trajectory for large field values
and will lead to different equation of state histories,
and hence different amounts of expansion.
$N(\theta)$ is shown in Figure~2,
where we integrated the equation of motion
starting from a fixed small initial value of $\phi$
and continuing until a fixed final energy density
where the equations of state had converged
\footnote{Here we ignore complications present in more complete models
where one has to take into account degrees of freedom suppressed here.
In particular, one might expect the different trajectories to lead to
different fractions of radiation or other things
produced at the end of the modular inflation.
We will discuss this in more detail in Section~\ref{thermalinf}.}.

{
\begin{figure}[ht]
\epsfxsize=0.48\textwidth
\begin{center}    
\leavevmode       
\epsffile{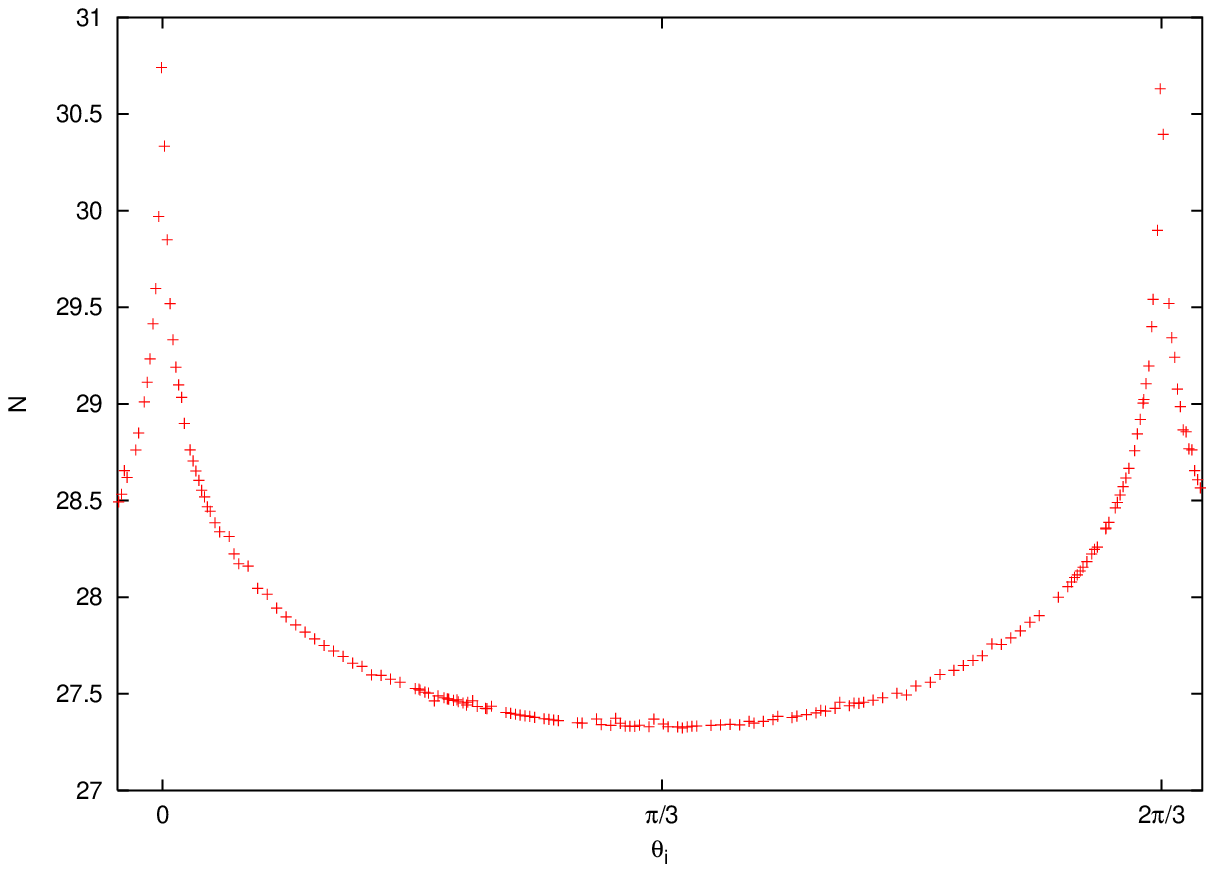}
\end{center}        
\caption{Number of $e$-folds $N$ as a function of
the initial angle $\theta_\mathrm{i}$.}
\label{fig2}
\end{figure}
}

We see that
\be\label{unity}
\frac{\partial N}{\partial \theta} \sim {\cal O}(1)
\ee
for a wide range of $\theta_i$,
though the initial values $\theta_i \sim 2n\pi/3$,
which lead to large spikes in $N$, could be a postiori preferred
due to the greater spatial volume produced \cite{ewankenji}.

Thus, from  Eqs.~(\ref{rad2}) and~(\ref{ang2}),
we see that the angular fluctuations give a comparable,
or possibly greater, contribution to the final curvature
perturbation.
This is due to the huge magnification of the angular fluctuations
as $\phi$ rolls from small values to Planckian values,
comparable to the usual huge magnification of radial fluctuations
near a maximum due to the factor
$\partial N/\partial \phi = - H/\dot{\phi}$.

However, as ${\partial N}/ \partial \theta$ does not depend
on the initial value of $\phi$ (assuming that it is small),
${\partial N}/ \partial \theta$ will be independent of $k$.
Thus the contribution of the angular fluctuations,
Eq.~(\ref{ang2}), has the same steep spectrum as that
of the radial fluctuations and is not consistent with observations.

In the next sections we introduce models which avoid this problem.

\subsection{Single Modulus with Quantum Corrections}
\label{singlecorrections}

Here we consider the next simplest model of modular inflation
given in Ref.~\cite{running}.
We first briefly describe the model and
then discuss the perturbations produced by it.

The previous model considered a modulus rolling away from a maximum
at a point of enhanced symmetry.
At such a point of enhanced symmetry,
one would in general expect the modulus to have couplings
to other light fields.
These couplings will cause the mass to renormalise as a function of
$\phi$.
This could turn the mass squared of the modulus positive for small
$\phi$, shifting the maximum of the potential out to some finite value
$\phi = \phi_*$.

Explicitly, following Ref.~\cite{running}
with the notation change $\epsilon \rightarrow \beta$ and
implicitly keeping the full complex modulus field $\Phi$,
\begin{equation}
V = V_0 \left[ 1 - \frac{1}{2}\,f(\beta\ln\phi)\,\phi^2 + \ldots \right]~,
\end{equation}
and we have a maximum at $\phi=\phi_*$ where
\begin{equation}
f_* + \frac{1}{2} \beta f'_* = 0~.
\end{equation}
We resolve the ambiguity in the definitions of $f$ and $\beta$ by fixing
$f_* = f(-1)$ so that $\phi_* = e^{-1/\beta}$.
Note that $f_* < 0$ and typically $\beta \ll 1$.
We assume $H \simeq \sqrt{V_0/3}$ and $\phi \simeq \phi_*$
when observable scales leave the horizon.

One will again have eternal inflation at the maximum,
which now corresponds to the ring $\phi=\phi_*$,
and the eternal inflation ends locally
when $\phi$ locally rolls off the maximum towards large values.
The distribution of $\theta$ values during the eternal inflation
could be biased by the small symmetry breaking effects of
the $\Phi^3$ terms, which should also be included in this model,
because the difference in expansion rate around the ring
leads to differences in the spatial volumes generated.
However, for small $\phi_*$, this biasing will be negligible
as the differential expansion rate is much smaller than
the diffusion rate around the ring due to quantum fluctuations.

This model has the interesting property that
in the neighborhood of the maximum the potential
is flattened in the
radial direction by a factor $\beta$.
This helps to flatten the contribution of the radial fluctuations
to the final curvature perturbation, but not by as much as one would
like \cite{running}.
However, we will be more interested in the effect of
the shift of the maximum out to finite $\phi=\phi_*$.
This means that as $\phi$ rolls off the maximum,
while $\phi-\phi_* \ll \phi_*$, the radius $\phi$ will hardly change,
allowing the angular fluctuations to give a flat contribution
to the final curvature perturbation.

Explicitly, the contribution of the radial fluctuations
to final curvature perturbation is 
\be
\frac{\partial N}{\partial\phi} \, \delta\phi
= - \frac{H}{\dot{\phi}} \, \delta\phi
\sim \frac{H}{\phi-\phi_*}~,
\ee
and the contribution of the angular fluctuations is
\be
\frac{\partial N}{\partial\theta} \, \delta\theta
\sim \frac{\partial N}{\partial\theta} \frac{H}{\phi}~.
\ee
Therefore, for $\partial N / \partial\theta \sim \mathcal{O}(1)$, 
the contribution of the angular fluctuations is suppressed
relative to that of the radial fluctuations by the factor
$(\phi-\phi_*)/\phi$.
This is easily understood as for large $\phi$,
\mbox{i.e.} on small scales, the contributions are comparable,
but as one approaches $\phi_*$, \mbox{i.e.} going to larger scales,
the angular contribution to the spectrum flattens out
while the radial contribution continues rising and so dominates.

However, if $\partial N / \partial\theta \gg 1$,
then the contribution of the angular fluctuations can dominate
the final curvature perturbation giving the desired flat spectrum.
\footnote{A rough estimate suggests we need
$\partial N / \partial\theta \gtrsim 10^3$.
See Appendix~\ref{dNdtheta}.}
This is realized, as indicated from Figure~\ref{fig2},
if the angle is close to the one of angular maxima of the potential
at $\theta = 2n\pi/3$.
As mentioned before, this can be probable if we consider that 
regions corresponding to these initial angles will undergo greater
expansion and hence occupy a larger volume at late times.
However, we do not want to be drawn too close to these special angles
as $\partial N / \partial\theta$ will either diverge or be zero there.
This requires further investigation \cite{ewankenji}.

A more precise formula for the spectrum of $\delta\theta$ is given by
Eqs.~(\ref{Psinglecase}) and~(\ref{singlecase}) in the Appendix.
The spectral index has the form
\be
\label{singlen}
n-1 = - A k^{\beta f'_*}~,
\ee
and the running
\be
\frac{d n}{d\ln k} = - A \beta f'_* k^{\beta f'_*}~.
\ee
Therefore, if the contribution of the angular fluctuations dominates,
one can expect a flat spectrum on larger scales
with the spectrum dipping away with substantial running on smaller scales.
It is very plausible that observable scales correspond to
$\phi$ close to, but not very close to, $\phi_*$, so that
this running would be observable \cite{WMAP}.

\subsection{A Second Modulus}
\label{secondcorrections}
String theory does not predict a single modulus, but rather many moduli.
Here, in addition to $\Phi$, we consider the case that there is a second modulus $\Psi$ which drives eternal modular inflation and also some post-eternal modular inflation, before $\Phi$ rolls down driving its own modular inflation.

We take $\Psi$'s potential to have the form
\be
V = V_0 - \frac{1}{2} m_\psi^2 \psi^2 + \ldots~,
\ee
where we consider just a single real direction $\psi \in \Psi$,
and, as always, $m_\psi^2 \sim V_0$ and everything is at
the vacuum supersymmetry breaking scale.
As before, $\psi$ rolls down as
\be
\psi \propto a^\alpha~,
\ee
where
\be
\alpha = \frac{3}{2}
\left( \sqrt{ 1 + \frac{4 m_\psi^2}{3 V_0} }\, - 1 \right)
\ee
and the contribution of fluctuations in $\psi$
to the final curvature perturbation will be
\be\label{psipert}
\frac{\partial N}{\partial\psi} \, \delta\psi
= - \frac{H}{\dot{\psi}} \, \delta\psi
\sim \frac{H}{\psi}~.
\ee

During the modular inflation driven by $\psi$,
$\Phi$'s potential will have the form
\be
V = V(\psi) + \frac{1}{2} m_\phi^2(\phi,\psi) \, \phi^2 + \ldots~.
\ee
The mass squared $m_\phi^2(\phi,\psi)$ is renormalised
as a function of $\phi$ as before.
However, it will also have a dependence on $\psi$ due to
the supersymmetry breaking effect of $\psi$'s potential.
Schematically, it looks like
\be
m_\phi^2(\phi,\psi) \sim - m_\mathrm{vac}^2 + m_\mathrm{inf}^2
+ \beta m_\mathrm{ren}^2 \ln\phi~,
\ee
where $m_\mathrm{vac}^2 \sim m_\mathrm{inf}^2
\sim m_\mathrm{ren}^2 \sim m_\psi^2$ and $\beta \ll 1$.
We assume the indicated signs and that $m_\mathrm{inf}^2 > m_\mathrm{vac}^2$
so that $m_\phi^2 > 0$ at large $\phi$ during $\psi$'s modular inflation.
We further assume that the renormalisation group running eventually
turns $m_\phi^2$ negative so that $\phi$ has a minimum
at some exponentially small value $\phi_*$.
This is just the inverse of the running we had in the previous model
which produced a maximum,
and leads to a quite interesting scenario as follows.

During $\psi$'s modular inflation,
$\phi$ has a mass of order $H$ which will suppress its fluctuations
\footnote{Note that the flattening mechanism also applies in this case,
so the fluctuations in $\phi$ may not be completely suppressed.
We will leave further investigation of this point to another paper
\cite{ewankenji}.},
while $\Phi$'s angular component stays massless
and so will have unsuppressed fluctuations.
Once $\psi$'s modular inflation ends, $m_\mathrm{inf}^2$ will disappear,
or at least change greatly, releasing $\phi$ to roll down to its
vacuum at Planckian values as in our previous models,
driving some more modular inflation as it does so.
The angular fluctuations in $\Phi$ are massively amplified by this roll down
and give a contribution to the final curvature perturbation
\be
\frac{\partial N}{\partial\theta} \, \delta\theta
\sim \frac{\partial N}{\partial\theta} \frac{H}{\phi_*}
\ee
assuming that observable scales leave the horizon
during $\psi$'s modular inflation.
Since $\partial N / \partial\theta$ is independent of $k$,
and $\phi_*$ is a constant, we get a flat contribution
to the final curvature perturbation.
Furthermore, it will dominate over $\psi$'s contribution as long as
$\psi \gg \phi_*$ when observable scales leave the horizon.
Finally, for $H \sim 10^{-16}$ and
$\partial N / \partial\theta \sim \mathcal{O}(1)$,
our model matches the COBE normalisation with a very reasonable
$\phi_* \sim 10^{-12}$, which is a natural consequence of
the logarithmic renormalisation group running in our weak coupling regime.

A more precise formula for the spectrum of $\delta\theta$
is given by Eqs.~(\ref{Presult}) and~(\ref{result}) in the Appendix.
The spectral index has the form
\be
\label{doublen}
n-1 = - A k^{2\alpha}~,
\ee
and the running
\be
\frac{d n}{d\ln k} = - 2 \alpha A k^{2\alpha}~.
\ee
So, as in our previous model, one can expect a flat spectrum on larger scales
with the deviation from a flat spectrum on smaller scales. 
In this model one expects $\alpha \sim 1$ so the running will typically be
larger than in the previous model.
Due to the second modular inflation driven by $\phi$,
it is very plausible that observable scales left the horizon
near the end of the first modular inflation driven by $\psi$,
so that this running may be observable \cite{WMAP}.

As mentioned in the Appendix, the above formula for the spectral index
applies only for $\alpha<1$.
The case $\alpha \gtrsim 1$ is more involved because
the deviation from scale invariance is dominated by non-trivial dynamics
at the end of inflation. 
Thus, in our model, even though the spectrum will be rather flat on large scales,
we could possibly expect deviation from a flat spectrum for smaller scales
caused by late-time inflation dynamics affecting super-horizon modes \cite{misao}.
We leave the detailed analysis of this perturbation calculation
on super-horizon scales during the late epoch of modular inflation
for our forthcoming paper \cite{ewankenji}.

\section{Thermal Inflation}
\label{thermalinf}

After our modular inflation, the moduli roll down and oscillate about
the minimum of their potential.
If this minimum corresponds to a point of enhanced symmetry,
where the modulus has unsuppressed couplings to other light fields,
then the modulus will decay rapidly.
If the minimum is a Planckian distance away from any such point,
then the modulus will have only gravitational strength couplings
to other light fields and will decay too late for the successful predictions
of Big Bang nucleosynthesis to survive \cite{Polonyi}.
Even if our modulus decays rapidly,
one expects other moduli to be displaced during the modular inflation
and to roll down and oscillate when it ends.
Again, these moduli may either decay rapidly,
or contribute to the Polonyi problem.
Thus, after the modular inflation, one expects to have initially
perhaps comparable fractions of radiation and dangerous long-lived moduli.

Thermal inflation is considered to be one of the most
promising mechanisms to get rid of the unwanted moduli.
We keep the following discussion of thermal inflation fairly brief
and refer the reader to Ref.~\cite{lythewan} for more details.

Thermal inflation is driven by a flaton $\sigma$ with potential of the form
\be
V = V_0 + \left( g^2 T^2 - m^2 \right) |\sigma|^2
+ \left( A \lambda \sigma^{n} + \mbox{h.c.} \right)
+ |\lambda|^2 |\sigma|^{2n-2}~,
\ee
with $m \sim 10^2-10^3$ GeV and $n = 4$ or $5$.
The flaton's vacuum expectation value is $M \sim m^{1/(n-2)}$
and $V_0 \sim m^2 M^2 \ll m^2$
so we do not expect eternal or modular inflation.
When $T > T_\mathrm{c} = m / g$,
the flaton is trapped at the origin by the finite temperature potential,
and inflation begins when $V_0$ dominates over the moduli density.
After about 5 to 10 $e$-folds of inflation, which dilutes the moduli,
the temperature drops below $T_\mathrm{c}$
and the flaton rolls rapidly towards its minimum ending the inflation.
The Universe is then dominated by the oscillating flaton,
which eventually decays to ordinary fields
through non-renormalisable operators suppressed by of order $M$.
After this `reheating' at temperatures of order GeV,
standard Big Bang cosmology follows.

We point out here that fluctuations in the angular part of the modulus
produced during the modular inflation, which lead to perturbed
trajectories at the end of the modular inflation, could lead to
perturbations in the fraction of radiation produced at the end of
the modular inflation.
This could then lead to perturbations in the amount of thermal inflation,
as the thermal inflation begins when the moduli density drops below $V_0$
but ends when the temperature drops below $T_\mathrm{c}$,
giving an extra contribution to $\partial N / \partial\theta$.
However, the perturbation in the fraction of radiation produced
at the end of the modular inflation is likely to be highly model
dependent and difficult to evaluate.
This will be further investigated in Ref.~\cite{ewankenji},
though we do not expect it to change our conclusions greatly.

We also note in Appendix~\ref{A3} that thermal fluctuations
can produce perturbations during thermal inflation,
but these are negligible in our case.

Thus thermal inflation, which is a natural consequence of thermal effects
in the presence of supersymmetric flat directions,
resolves our remaining difficulty with modular cosmology.

\section{Conclusions}
\label{dis}

We presented the first realistic modular cosmology scenario
in that dilaton stabilization, inflation producing a scale invariant
spectrum of density perturbations, and Polonyi/moduli problems
are resolved in a simple and self-consistent manner.

One of the notable features is eternal inflation
at the vacuum supersymmetry breaking scale.
This causes the fine-tuned initial conditions required
for the dilaton to find its way to a weak coupling minimum
to become a postiori as natural as any others
because of the subsequent eternal inflation.

Our model also predicts the observed scale-invariant
cosmic microwave background spectrum without any fine-tuning,
and even allows for the possibility of a dipping spectrum
with significant running on smaller scales \cite{WMAP},
which alone could deserve special attention
from the viewpoint of inflation model building.
Indeed, non-slow-rolling inflaton fields are generic
in supergravity theory, but usually avoided by fine-tuning
because of their non-scale-invariant power spectrum.
This fine-tuning is not necessary in our scenario
because the density perturbation is dominantly produced from
the angular degree of freedom of a modulus whose usually subdominant
fluctuations are hugely amplified by the long roll out of the modulus.
We gave explicit calculations of the spectrum, spectral index
and its running for modular inflation for concrete toy models.
We also pointed out the possible 
deviation from scale invariance 
at small scales for some values of the parameters, 
and we leave thorough treatment of this issue 
for our forthcoming paper \cite{ewankenji}.

Finally, thermal inflation solves the Polonyi/moduli problem.

Due to the generic existence of ubiquitous moduli in string/M-theory,
an early universe history where moduli play crucial roles
would be a rather natural consequence.
We kept our discussions fairly general
so that it can be applied to a wide class of models,
and hope our proposed scenario leaves room for
the further investigation of successful modular cosmology.

\subsection*{Acknowledgements}    
We thank Joanne D. Cohn for catalyzing our collaboration and 
for encouragement,
and the SF02 Cosmology Summer Workshop for hospitality
when this collaboration began.
E.D.S. thanks the Fermilab Theoretical Astrophysics Group
for hospitality while this work was in progress.
K.K. thanks J.D. Cohn and H. Murayama for their continuous encouragements. 
E.D.S. was supported in part by
the Astrophysical Research Center for the Structure and Evolution
of the Cosmos funded by the Korea Science and Engineering Foundation
and the Korean Ministry of Science,
and by the Korea Research Foundation grants KRF-2000-015-DP0080 and KRF PBRG
2002-070-C00022.
K.K. was supported in part by NSF under grant AST-0205935.

\setcounter{section}{0}
\renewcommand{\thesection}{\Alph{section}}
\section{Appendices}

\subsection{Precise Calculation of $\delta\theta$}
\label{general}

In this appendix we calculate the spectra for the models presented in
Sections~\ref{singlecorrections} and~\ref{secondcorrections},
starting with a general formula applicable to both.

To calculate $\delta\theta$,
on superhorizon scales but while the angular potential is still flat,
we need to solve the perturbed equation of motion
for $\delta\theta(k,t)$
\begin{equation}
\label{perturbation}
\frac{d}{dt} \left( a^3 \phi^2 \frac{d\,\delta\theta}{dt} \right)
+ a k^2 \phi^2 \delta \theta = 0~.
\end{equation}
Defining
\begin{equation}
\varphi \equiv a \phi \, \delta\theta
\end{equation}
and using the conformal time $d\eta \equiv dt/a$, we obtain
\begin{equation}
\label{onemodulus}
\frac{d^2\varphi}{d\eta^2}
+ \left[ k^2 - \frac{1}{a\phi} \frac{d^2(a\phi)}{d\eta^2} \right] \varphi = 0~,
\end{equation}
where
\ba
\frac{1}{a\phi} \frac{d^2(a\phi)}{d\eta^2} =               
2 a^2 H^2 \left( 1 + \frac{1}{2 H^2} \frac{dH}{dt}
+ \frac{3}{2 H \phi} \frac{d\phi}{dt}
+ \frac{1}{2 H^2 \phi} \frac{d^2 \phi}{dt^2} \right)~.
\ea
Introducing $y \equiv \sqrt{2k}\, \varphi$ and $x \equiv -k \eta$,
we can write this as
\be
\label{ansatzeqn}
\frac{d^2 y}{d x^2} + \left( 1 - \frac{2}{x^2} \right) y
= \frac{g}{x^2} y~,
\ee
where
\be
g = \frac{1}{x a \phi} \left[ \frac{d^2(x a \phi)}{(d\ln x)^2}
- 3 \frac{d(x a \phi)}{d\ln x} \right]~.
\ee
For
\be
g = \xi \left(\frac{x}{x_\star}\right)^\nu
\ee
with $\xi \ll 1$, this can be solved to give \cite{newan}
\be
\label{finalP}
P_{\delta\theta}(k) = \left( \frac{k}{2\pi x_\star a_\star \phi_\star}\right)^2
\left\{ 1 + \frac{2}{(3-\nu)\nu}
\left[ (2 x_\star)^{-\nu} \cos\left(\frac{\pi\nu}{2}\right)
\frac{\Gamma(2+\nu)}{1-\nu} - 1 \right] \xi
+ {\cal O}(\xi^2) \right\}~,
\ee
where subscript $\star$ denotes an appropriate evaluation point.
The divergence at $\nu = -2$ corresponds to
the deviation of the spectrum from scale invariance being dominated by
late time dynamics \cite{misao}.
The spectral index is
\be
n_{\delta\theta}(k) - 1
= - \cos\left(\frac{\pi\nu}{2}\right)
\frac{2\Gamma(2+\nu)}{(1-\nu)(3-\nu)}
\frac{\xi}{(2 x_\star)^\nu}~.
\ee
We apply the above formula to the specific models we discussed in
the paper.

\subsubsection{Single Modulus with Quantum Corrections}
\label{A1}

Here we derive the explicit form for Eq.~(\ref{singlen})
of Section \ref{singlecorrections}.

For $H \simeq \sqrt{V_0/3}\,$, we have
\be
x = \frac{k}{aH}~,
\ee
and
\be
g = \frac{3}{\phi H} \frac{d\phi}{dt}
+ \frac{1}{\phi H^2} \frac{d^2 \phi}{dt^2}~.
\ee
Eq.~(24) of Ref.~\cite{running} gives
\begin{equation}
\ln\left(\frac{\phi}{\phi_*}\right) = \frac{1}{\beta} e^{- \beta f'_* (N+C)}~,
\end{equation}
where $N$ is the number of $e$-folds until inflation ends at $\phi \sim 1$,
and $C \sim 1/\beta$.
Therefore
\begin{equation}
\frac{1}{\phi H} \frac{d\phi}{dt} = f'_* e^{- \beta f'_* (N+C)}
\end{equation}
and
\be
\frac{1}{\phi H^2} \frac{d^2 \phi}{dt^2}
= f'_* e^{- \beta f'_* (N+C)}
\left[ \beta f'_* + f'_* e^{- \beta f'_* (N+C)} \right]
\simeq \beta {f'_*}^2 e^{- \beta f'_* (N+C)}~.
\ee
Hence
\begin{equation}
g = \left( 3 + \beta f'_* \right) f'_* e^{- \beta f'_* (N+C)}
= \left( 3 + \beta f'_* \right) \beta f'_*
\ln\left(\frac{\phi_\star}{\phi_*}\right)
\left(\frac{x}{x_\star}\right)^{-\beta f'_*}~,
\end{equation}
leading to
\be\label{Psinglecase}
P_{\delta\theta}(k) = \left(\frac{H_\star}{2\pi \phi_\star}\right)^2
\left\{ 1 - 2 \left[ 2^{\beta f'_*} \cos\left(\frac{\pi\beta f'_*}{2}\right)
\frac{\Gamma(2 - \beta f'_*)}{1 + \beta f'_*}
\left(\frac{k}{a_\star H_\star}\right)^{\beta f'_*} - 1 \right]
\ln\left(\frac{\phi_\star}{\phi_*}\right) \right\}
\ee
and
\be
\label{singlecase}
n_{\delta\theta}(k) - 1
= - 2^{\beta f'_*} \cos\left(\frac{\pi\beta f'_*}{2}\right)
\frac{\Gamma(2-\beta f'_*)}{(1+\beta f'_*)}
2 \beta f'_* \ln\left(\frac{\phi_\star}{\phi_*}\right)
\left(\frac{k}{a_\star H_\star}\right)^{\beta f'_*}~.
\ee
Note the sign change from one's naive expectation for $\beta f'_* > 1$
\cite{ewankenji}.

\subsubsection{A Second Modulus}
\label{A2}

Now we derive the explicit form for Eq.~(\ref{doublen})
of Section~\ref{secondcorrections}.

Here we assume $\phi=\phi_*$ leaving consideration of the effects of any time
dependence of $\phi$ to another paper \cite{ewankenji}. $\psi$ scales as 
\be
\psi \propto a^\alpha
\ee
with
\be
\alpha = \frac{3}{2}
\left( \sqrt{ 1 + \frac{4 m_\psi^2}{3 V_0} }\, - 1 \right)~.
\ee
Therefore
\be
3 H^2 = V_0 - \frac{1}{2} m_\psi^2 \psi^2 + \frac{1}{2} \dot{\psi}^2
= V_0 \left( 1 - \frac{1}{2} \alpha \psi^2 \right)
\label{ep}
\ee
and
\be
x = \frac{k}{aH} \left[ 1 + \frac{\alpha^2 \psi^2}{2(1-2\alpha)} \right]~.
\ee
Hence
\be\label{aa}
g = \frac{1}{2} \left(\frac{3+2\alpha}{1-2 \alpha}\right)
\alpha^2 \psi_\star^2 \left(\frac{x}{x_\star}\right)^{-2\alpha}~,
\ee
leading to
\be
P_{\delta\theta}(k) =
\left(\frac{H_\star}{2\pi\phi_*}\right)^2
\left\{ 1 - \frac{\alpha}{2}
\left[ 2^{2\alpha} \cos(\pi\alpha)
\frac{\Gamma(2-2\alpha)}{1-4\alpha^2}
\left(\frac{k}{a_\star H_\star}\right)^{2\alpha} - 1 \right]
\psi_\star^2 \right\}
\label{Presult}
\ee
and 
\be
n_{\delta\theta}(k)-1 =
- 2^{2\alpha} \cos(\pi\alpha) \frac{\Gamma(2-2\alpha)}{1-4\alpha^2}
\alpha^2\psi_\star^2 \left(\frac{k}{a_\star H_\star}\right)^{2\alpha}~.
\label{result}
\ee

\subsection{Estimation of the Required Value of
$\partial N / \partial \theta$ for the
Single Modulus with Quantum Corrections Model}
\label{dNdtheta}

In the single modulus with quantum corrections model
of Section~\ref{singlecorrections},
the radial fluctuations have only a slightly flattened spectrum.
Therefore we require
\be
\frac{\partial N}{\partial \phi} \, \delta\phi
\lesssim 10^{-1} \frac{\partial N}{\partial \theta} \, \delta\theta  
\ee
to avoid the radial fluctuations giving the full spectrum
too big a deviation from scale invariance.
Now
\be
\delta\theta \sim \frac{\dot\phi}{H\phi}
\frac{\partial N}{\partial\phi} \, \delta\phi
\ee
and using the results of Appendix~\ref{A1} we have
\be
\frac{\dot\phi}{H\phi} \sim \beta f'_* \ln\left(\frac{\phi}{\phi_*}\right)
\sim \frac{n_{\delta\theta}-1}{2}
\lesssim 10^{-2}~.
\ee
Thus we get the estimate
\be
\frac{\partial N}{\partial\theta} \gtrsim 10^3~.
\ee

We also note that matching the COBE normalization requires
\be
\frac{\partial N}{\partial\theta} \frac{H}{\phi_*} \sim 10^{-4}~,
\ee
and so for $H \sim 10^{-16}$ require
\be
\frac{\partial N}{\partial\theta} \sim 10^{12} \phi_*
\ee
which gives us an upper limit on $\partial N / \partial\theta$
as we expect $\phi_*$ to be well below the Planck scale.

\subsection{Perturbations from Thermal Fluctuations during Thermal Inflation}
\label{A3}

In this appendix we note that
there can arise another source of perturbations 
caused by thermal fluctuations during thermal inflation.
These thermal fluctuations can be estimated by considering 
the thermal fluctuation $\delta T \sim T$
in regions of the thermal bath of size $\sim 1/T$.
A Hubble volume of radius $1/H$,
outside of which the thermal fluctuation freezes in,
encloses $T^3/H^3$ of such regions
giving a thermal fluctuation on the Hubble scale of
$\delta T \sim T / \sqrt{T^3/H^3}$.
Thus we have
\be
\mathcal{R}_\mathrm{c} = \delta N
= \frac{\partial N}{\partial T} \, \delta T
\sim \frac{1}{T} \cdot T \sqrt{\frac{H^3}{T^3}}
\sim \left(\frac{T_\mathrm{c} M}{T}\right)^{3/2}~.
\ee
Now $T \propto a^{-1}$ and so this has a spectral index $n=4$,
with its maximum value $\sim M^{3/2}$ at the end of thermal inflation.
This is completely negligible for thermal inflation
with an intermediate scale $M$ as we are considering,
but could be significant for thermal inflation
driven by a flaton with $M \gtrsim M_\mathrm{GUT}$.


\end{document}